\documentclass{Interspeech2024}

\interspeechcameraready

\title{Speech foundation models in healthcare: Effect of layer selection on pathological speech feature prediction}

\name[affiliation={1}]{Daniela A.}{Wiepert}
\name[affiliation={1}]{Rene L.}{Utianski}
\name[affiliation={1}]{Joseph R.}{Duffy}
\name[affiliation={1}]{John L.}{Stricker}
\name[affiliation={1}]{Leland R.}{Barnard}
\name[affiliation={1}]{David T.}{Jones}
\name[affiliation={1}]{Hugo}{Botha}

\address{
 $^1$Department of Neurology, Mayo Clinic (Rochester), USA}
\email{wiepert.daniela@mayo.edu, botha.hugo@mayo.edu}


\begin{document}

\maketitle

\begin{abstract}
    
    Accurately extracting clinical information from speech is critical to the diagnosis and treatment of many neurological conditions. As such, there is interest in leveraging AI for automatic, objective assessments of clinical speech to facilitate diagnosis and treatment of speech disorders. We explore transfer learning using foundation models, focusing on the impact of layer selection for the downstream task of predicting pathological speech features. We find that selecting an optimal layer can greatly improve performance ($\sim$15.8\% increase in balanced accuracy per feature as compared to worst layer, $\sim$13.6\% increase as compared to final layer), though the best layer varies by predicted feature and does not always generalize well to unseen data. A learned weighted sum offers comparable performance to the average best layer in-distribution (only $\sim$1.2\% lower) and had strong generalization for out-of-distribution data (only 1.5\% lower than the average best layer). 
\end{abstract}

\section{Introduction}
Accurately extracting clinical information from speech is critical to the diagnosis and treatment of many neurological conditions that directly impact speech production. In practice, a speech pathologist or neurologist will conduct a perceptual assessment to examine whether pathological speech features are present in speech. This is a highly specialized field with a global shortage of access to expertise especially in rural and developing regions. Many are unable to be seen by clinicians with the skills to conduct such assessments and diagnose accordingly, resulting in preventable morbidity \cite{Dang21}. Automated or semi-automated tools powered by artificial intelligence (AI) offer one means to address this gap. Such tools may be able to conduct automatic, objective assessments of clinical speech, in turn facilitating triage and referral or diagnosis and treatment by non-experts without access to specialists \cite{Dorsey13, Koch10}. 

Previous research has shown promise for creating diagnostic tools, but current speech models do not generalize well enough to be used in clinical practice \cite{Gupta16, Cohen21}. One challenge with the clinical application of speech models is that many diseases can impact speech, making it near impossible to have enough data to learn how to directly predict a medical or neurologic diagnosis. A potential solution is to instead focus on predicting the pathological speech features that characterize speech disorders \cite{Soltau23} as there are predictable mappings between pathological features and disorders \cite{Duffy19, Tripathi21}. These features are also not unique with respect to disease \cite{Duffy19, Tripathi21}. A model could therefore learn the information necessary to recognize a specific type of dysarthria using recordings from a cohort with varying neurological diseases or abnormalities, requiring less data overall. 

There is still not enough diverse, publicly shared clinical speech data to build such models from scratch \cite{Christensen12, Gupta16, Le18, Tripathi21}, likely because of privacy concerns with sharing clinical speech recordings \cite{Wiepert24} and the cost required to acquire and annotate large clinical datasets. This has led to increased interest in exploring transfer learning using foundation models. However, existing speech foundation models are pre-trained with healthy speakers for non-clinical tasks (e.g., ASR, speaker verification), meaning the resulting speech representations may not be optimal for clinical applications. Recent work has shown that using intermediate representations may offer performance improvements for downstream tasks outside of the pre-training objective as different model layers encode different speech information \cite{Pasad21, Pasad23, Chowdhury24}. It is possible that some layers may capture more clinically relevant speech information, leading to large performance improvements for downstream clinical tasks with minimal fine-tuning of the foundation models. 

In this work, we explore how speech representations extracted from different layers of a foundation model (\textit{wav2vec 2.0}) influence the performance and generalizability of classifiers trained on a novel clinical dataset to detect fine-grained pathological speech features which characterize various neurologic diseases. We compare the impact of layer selection to that of standard hyperparameter tuning and architecture alterations and additionally consider whether a learnable weighted sum combining speech representations from all layers \cite{Pasad21, Pasad23} is a suitable alternative when predicting multiple features.

\section{Related Work}
Previous work exploring automated or semi-automated AI tools for diagnosis of speech pathologies usually attempts to separate single diseases from healthy speakers or classify a handful of diseases \cite{Bertini21, Quan22, Wagner23}. These models tend to show good performance ($>80\%$ accuracy) \cite{Bertini21, Quan22} but are subject to methodological weaknesses. Binary classification of disease is not generally useful in clinical applications, especially when multiple diseases can present with similar underlying pathological speech features \cite{Duffy19}. Furthermore, most previous models have been trained with small, highly curated datasets, which calls their generalizability into question \cite{Bertini21, Quan22}. 

A more recent approach to working with existing speech data has been to leverage speech foundation models through transfer learning, both for clinical applications \cite{Wagner23} and other domains \cite{Li23}. One recent study found that \textit{wav2vec 2.0} representations are well-suited to encode characteristics of various speech pathologies \cite{Wagner23}. Another study showed that fine-tuning a pre-trained model to predict pathological speech features out-performed models trained from scratch \cite{Soltau23}. This study also found that representations from the middle layers of the model had higher prediction accuracy \cite{Soltau23}. They theorized that this was due to those layers providing a mix of acoustic and phonetic information that better encoded pathological features \cite{Soltau23}. Similar layer effects were shown in emotion recognition studies \cite{Li23}. These results align well with recent layer-analysis work examining where phone and word information concentrates in speech foundation models \cite{Pasad21, Pasad23}. These studies found that models encode the information in different layers, with the location dependent on the original pre-training objective \cite{Pasad21, Pasad23}. Taken together, this suggests that selecting the optimal intermediate representation from a foundation model will be important when applying transfer learning to clinical downstream tasks.

\section{Methods}
\subsection{Data}
Our speech dataset consists of recordings of elicited speech tasks gathered from speech assessments conducted both at Mayo Clinic and remotely. Recordings were annotated by a speech pathologist to indicate the presence of a selection of pathological speech features. Annotators reached consensus on a subset of recordings and created standard guidelines as a method of quality control. Our study had IRB approval and each speaker was consented prior to recording. Due to the clinical nature of the samples, this dataset is not publicly available.

We focused on two elicited tasks with overlapping feature annotations — Alternating Motion Rates (AMR - "puh puh puh") and Sequential Motion Rates (SMR - "puh tuh kuh"), which both have annotations for rate of speech (slow, rapid), strained voice, distortions in speech sounds, and irregular articulatory breakdowns. These features vary in domain of impairment \cite{Duffy19}, which may translate to different optimal model layers.

The AMR recordings were split by speakers into a train set of 487 speakers (686 recordings), a validation set of 50 speakers (74 recordings), and a test set of 96 speakers (136 recordings). The SMR recordings were an out-of-distribution test set with 209 all new speakers (223 recordings). The pathological feature breakdown is shown in Table \ref{tab:feature}. The AMR recordings averaged at 6.25 seconds and SMR recordings averaged at 11.17 seconds. 

The dataset was majority white, non-Hispanic, American speakers ($>95\%$) who were middle aged or older ($75\% > 50$ years old, average age of 59). We had diagnostic information for 632 AMR speakers and 186 SMR speakers. In the AMR samples, 81.8\% of speakers had a motor speech disorder (MSD, e.g. Dysarthria, Apraxia of Speech), 13.1\% had an undifferentiated MSD (i.e., uncertain type), 22.2\% had non-MSD abnormalities (e.g., smoker’s voice), and 7.1\% had no abnormalities. In the SMR samples, 22\% had an MSD, 2.7\% had an undifferentiated MSD, 32.8\% had non-MSD abnormalities, and 55.9\% had no abnormalities. Note that speakers could have more than one diagnosis (e.g., a patient may have Apraxia of Speech and Dysarthria).

\begin{table}[th]
\caption{Pathological feature breakdown by recording. s = strained voice, iab = irregular articulatory breakdowns, rr = rapid rate, sr = slow rate, d = distortions.}
\label{tab:feature}
\centering
\begin{tabular}{l|lllll}
\toprule
\textbf{Task} & \textbf{s} & \textbf{iab} & \textbf{rr} & \textbf{sr} & \textbf{d} \\ \hline
AMR &  174 & 344 & 152 & 401 & 407 \\ \hline
SMR & 24 & 9 & 4 & 22 & 42 \\
\bottomrule
\end{tabular}
\end{table}

\subsection{Model}
\begin{figure}[t]
  \centering
  \includegraphics[width=\linewidth]{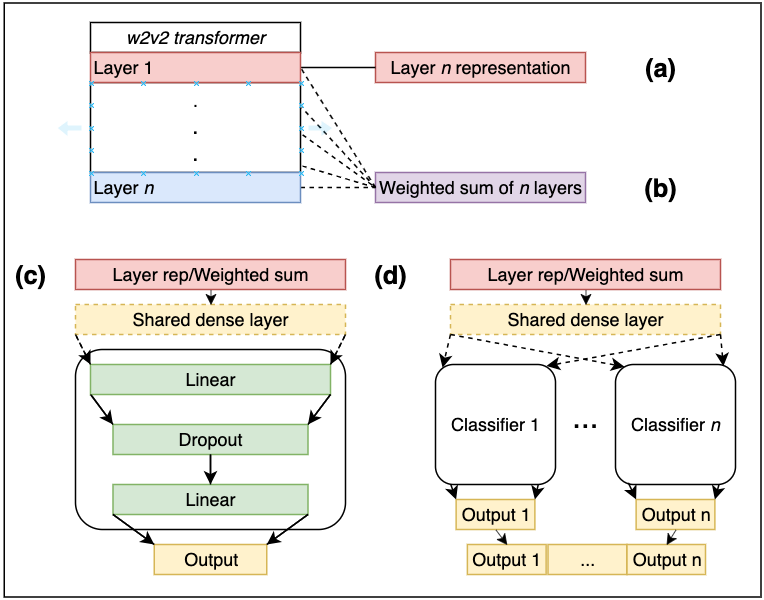}
  \caption{Overview of model architecture. To predict the presence of pathological speech features, we extracted either (a) a single layer from wav2vec 2.0 or (c) combined all layers in a learnable weighted sum then fed the resulting representations to (c) a single classifier or (d) multiple classifiers, with an optional shared dense layer prior to the classifier.}
  \label{fig:model}
\end{figure}

Using PyTorch \cite{pytorch}, we implemented a model that took processed speech recordings (16kHz, mono channel), fed them through a frozen \textit{wav2vec 2.0 base} model \cite{Baevski20, huggingface}, and used the intermediate representation(s) as input to a classifier. We selected \textit{wav2vec 2.0} due to its prior success in low resource domains \cite{Baevski20, Wiepert24} and its ability to learn discrete units of speech, which is useful in clinical settings where patients are often asked to make individual sounds (AMRs/SMRs). The base classifier includes a linear layer with ReLU activation, a dropout layer, and a final linear projection layer outputting predictions for each pathological speech feature (see Figure \ref{fig:model}). The source code is available at \href{https://github.com/MayoNeurologyAI/naip-w2v2/}{https://github.com/MayoNeurologyAI/naip-w2v2/}.

We compared the following classifier architectures:
 \begin{enumerate}
\item One classifier head to predict all pathological features (Figure \ref{fig:model}c) or separate classifier heads for each feature (Figure \ref{fig:model}d), to determine if parameter sharing helps classification.  

\item A shared dense layer with ReLU activation and an optional bottleneck parameter added prior to the classification head(s) (Figure \ref{fig:model}c-d) to assess if a learnable, shared representation prior to classification improves performance.  

\item A learnable weighted sum across layers \cite{Pasad21, Pasad23}, added immediately after the transformer block. This enables the model to use hidden representations from different transformer layers, which may improve performance (Figure \ref{fig:model}b). 
\end{enumerate}

The base single classifier configuration had $\sim$600,000 trainable parameters while the base multi-classifier configuration had $\sim$2.9M trainable parameters. Adding a shared dense layer added an additional $\sim$600,000 trainable parameters, and adding a learnable weighted sum added only 13 parameters.

The models were trained on 1 Nvidia A100 GPU for 20 epochs with batch size of 8 using only AMR recordings as input. The train, validation, and test sets were kept static across runs. We used the AdamW \cite{Adamw} optimizer and binary cross entropy loss. For reproducibility, we set a fixed manual seed of 4200. 

We conducted a grid search with the following hyperparameters: learning rate (1e-4, 1e-3), weight decay (1e-4, 1e-3, 1e-2), dropout probability (0.2, 0.3), classifier bottleneck (None/768, 700, 300), shared dense layer bottleneck (None/768, 700, 300). This search was run with the different combinations of classifier architectures across all layers (classifier head (2) x shared dense (2) x layer including weighted sum (14)). Of the hyperparameters, only learning rate had a large impact on performance. As such, we fixed the following parameters at their optimal values when presenting results: weight decay = 1e-4, dropout probability = 0.3, classifier bottleneck = None, shared dense layer bottleneck = None. 

We calculated the balanced accuracy to account for imbalance in the target classes as well as the no information rate to assess whether performance is better than chance. We additionally conducted bootstrap sampling of the predictions ($n=1000$) to calculate the 95\% confidence intervals. Each trained model was evaluated first with the AMR test set and then with the out-of-distribution SMR test set. We excluded predictions of irregular articulatory breakdowns and rapid rate in the out-of-distribution test due to low occurrences in the SMR set (see Table \ref{tab:feature}).

\section{Results}
\subsection{Hyperparameter tuning}
\begin{figure}[t]
  \centering
  \includegraphics[width=\linewidth]{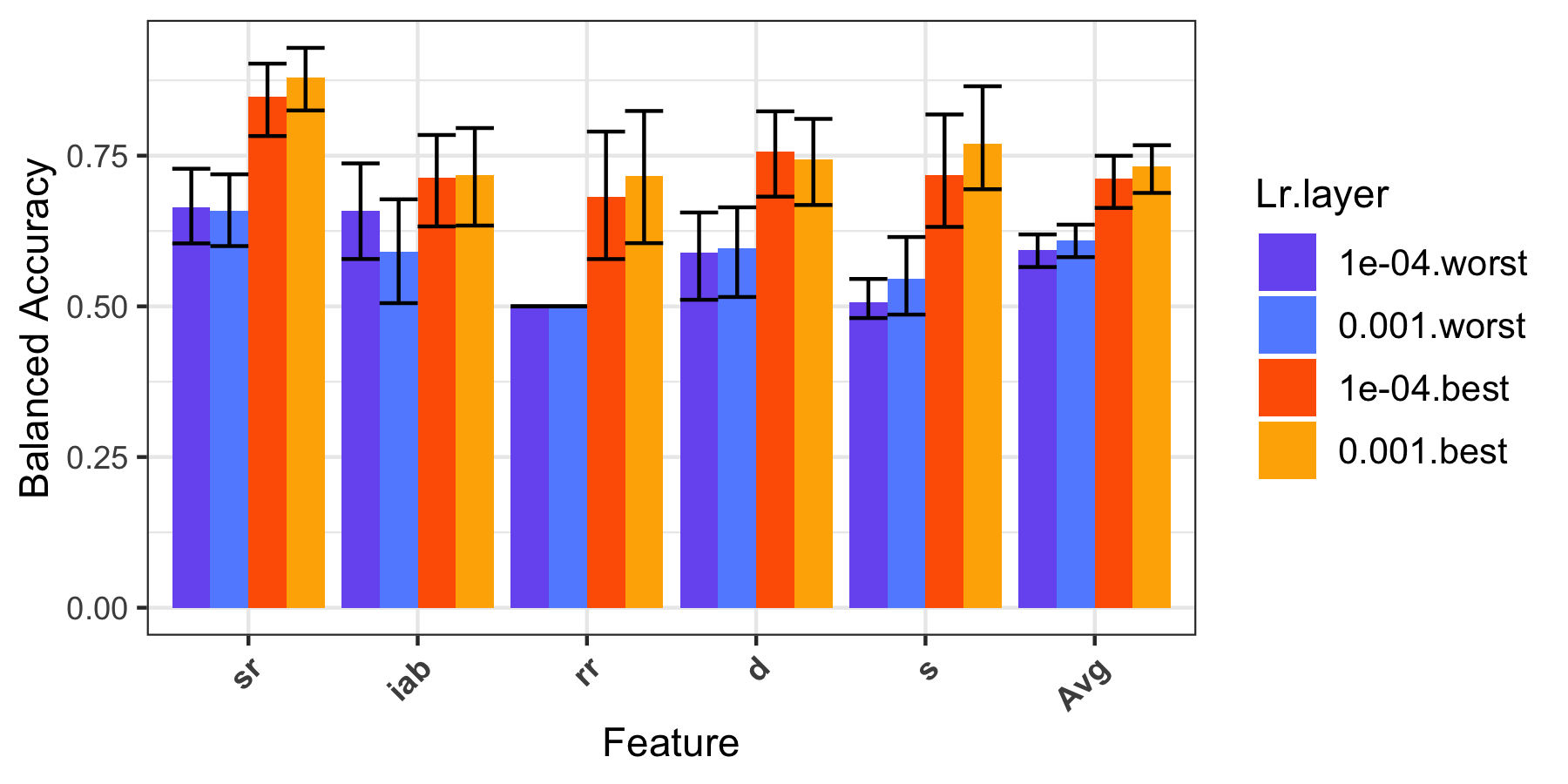}
  \caption{Comparing balanced accuracy by learning rate and layer across the predicted pathological speech features, with 95\% confidence intervals (AMR). }
  \label{fig:lr}
\end{figure}

The only hyperparameter that consistently impacted performance was the learning rate, with a learning rate of 1e-3 offering comparable or slightly better performance on the best layer for almost all features.  On average, this difference was small compared to the effect of layer selection (Figure \ref{fig:lr}, best vs. worst layer $= +16.5\%$, best learning rate $= +2.4\%$). These trends held for all model configurations. 

\subsection{Classifier architecture}

\begin{table}[th]
  \caption{Balanced accuracy averaged across features for each classifier architecture and layer (AMR).}
  \label{tab:bacc}
  \centering
  \begin{tabular}{lcccc}
  \hline
Layer & 1, sd & 1, No sd  &5, sd & 5, No sd \\ 
  \hline
  0 & 0.67 & 0.69 & 0.66 & 0.69 \\ 
  1 & \textbf{0.70}& 0.71 & 0.68 & 0.68  \\ 
  2 & 0.69 & 0.69 & 0.68 & 0.69 \\ 
  3 & 0.69& \textbf{0.72}& \textbf{0.71} & \textbf{0.70}   \\ 
  4 & 0.69& 0.70& 0.70 & 0.69   \\ 
  5 & 0.69& 0.71& 0.68 & 0.67   \\ 
  6 & 0.68  & \textbf{0.72}& 0.68 & 0.68 \\ 
  7 & 0.67 & 0.67 & 0.66 & \textbf{0.70}  \\ 
  8 & 0.66 & 0.69 & 0.65 & 0.68  \\ 
  9 & 0.67& 0.69 & 0.67 & 0.68 \\ 
  10 & 0.66 & 0.68 & 0.65 & 0.69  \\ 
  11 & 0.62 & 0.63 & 0.65 & 0.66  \\ 
  12 & 0.60& 0.61 & 0.60 & 0.61   \\ \hline
  Weighted Sum & 0.68 & 0.70 & 0.69 & 0.69 \\ 
   \hline
\end{tabular}
  
\end{table}

Changes in classifier architecture also had minimal impact on performance as compared to layer (Table \ref{tab:bacc}, best vs. worst layer $=+10.25\%$, best vs. worst architecture $=+3.1\%$). In general, the early to middle layers of the \textit{wav2vec 2.0} model offered much better performance across all classifier architectures. The final layer of the model was frequently the worst, an important point considering the most common approach with transfer learning using this model is to use the final layer. Notably, the learned weighted sum of layers offered performance comparable to the best layers (within 2\%) while adding very few trainable parameters. In general, the single classifier had slightly better performance ($+2\%$ for the best layers) while the addition of the shared dense layer never improved performance. The single classifier with no shared dense layer had the fewest trainable parameters of all configurations, meaning the difference in performance is not simply due to having more learnable parameters. 

\begin{figure}[t]
  \centering
  \includegraphics[width=\linewidth]{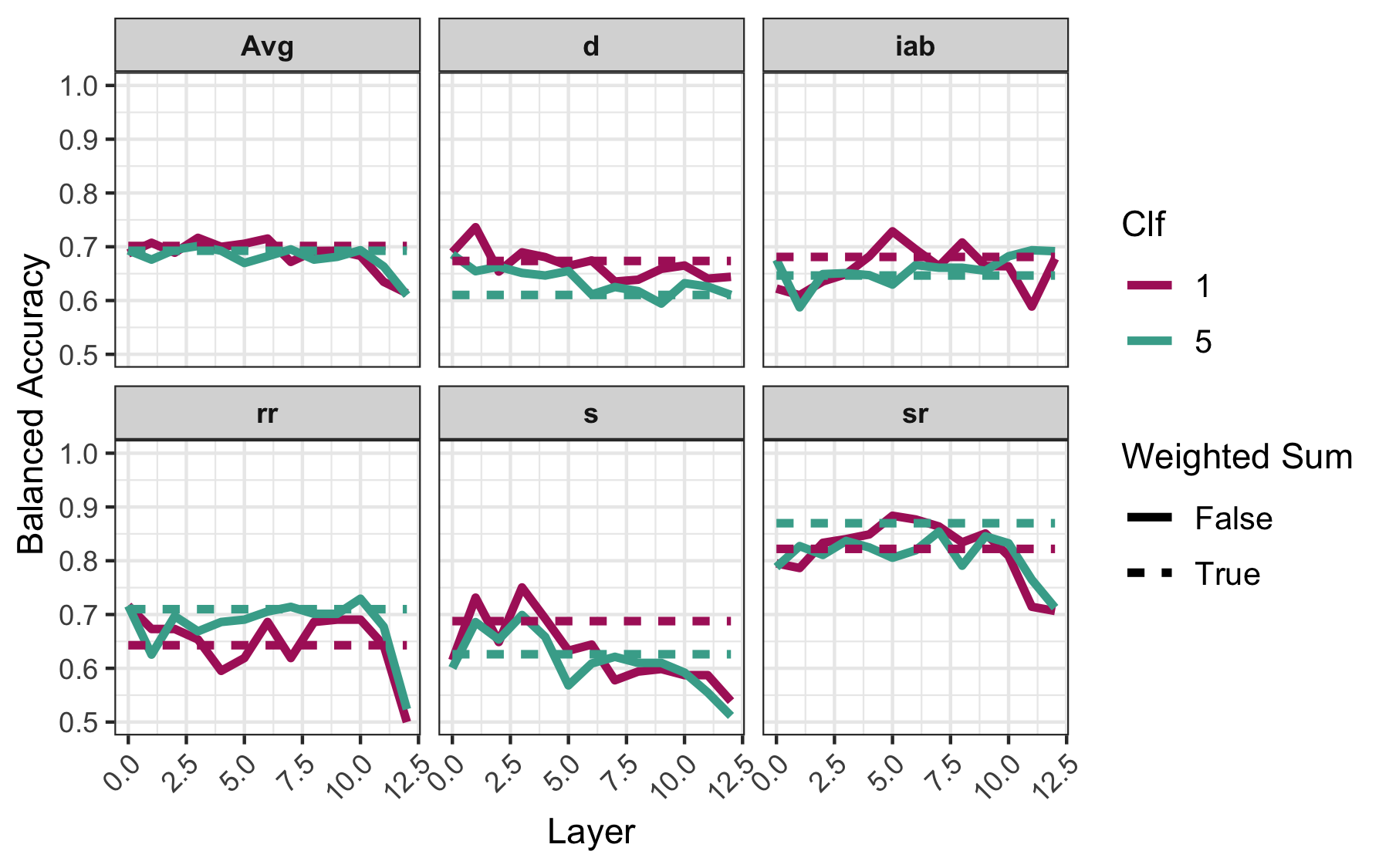}
  \caption{Comparing balanced accuracy across layers for each predicted pathological speech feature (AMR). }
  \label{fig:lineamr}
\end{figure}

When looking closer at the impact of layer across predicted features, we saw that the performance still tended to peak in earlier or middle layers regardless of feature, with the final layer often performing the worst (Figure \ref{fig:lineamr}). A single classifier also generally offered better performance than multiple classifiers. At the best layers, performance was significantly higher ($p<0.05$) than the no information rate for all features, but the same was not true for the worst layers, where only slow rate had any significant difference ($p<0.001$). Choosing the best layer per feature improved performance by an average of $\sim$15.8\% as compared to the worst layer and $\sim$13.6\% as compared to the final layer. The best layer varied across feature regardless of the classifier architecture (e.g., distortions peaked in layer 2 while slow rate peaked in layer 5). As a result, the average best layer was usually not the best layer for any given feature. The learnable weighted sum also surpassed the worst layer by a large margin of $\sim$10.7\% per feature, $\sim$8.5\% as compared to the final layer, but it was not as comparable to the best single layer per feature ($\sim$5.1\% lower than best layers). It was more comparable to the average best layer (only $\sim$1.2\% lower). The weighted sum performance was also somewhat significantly different ($p < 0.05$) from the no information rate for all features except strained.

\begin{figure}[t]
  \centering
  \includegraphics[width=\linewidth]{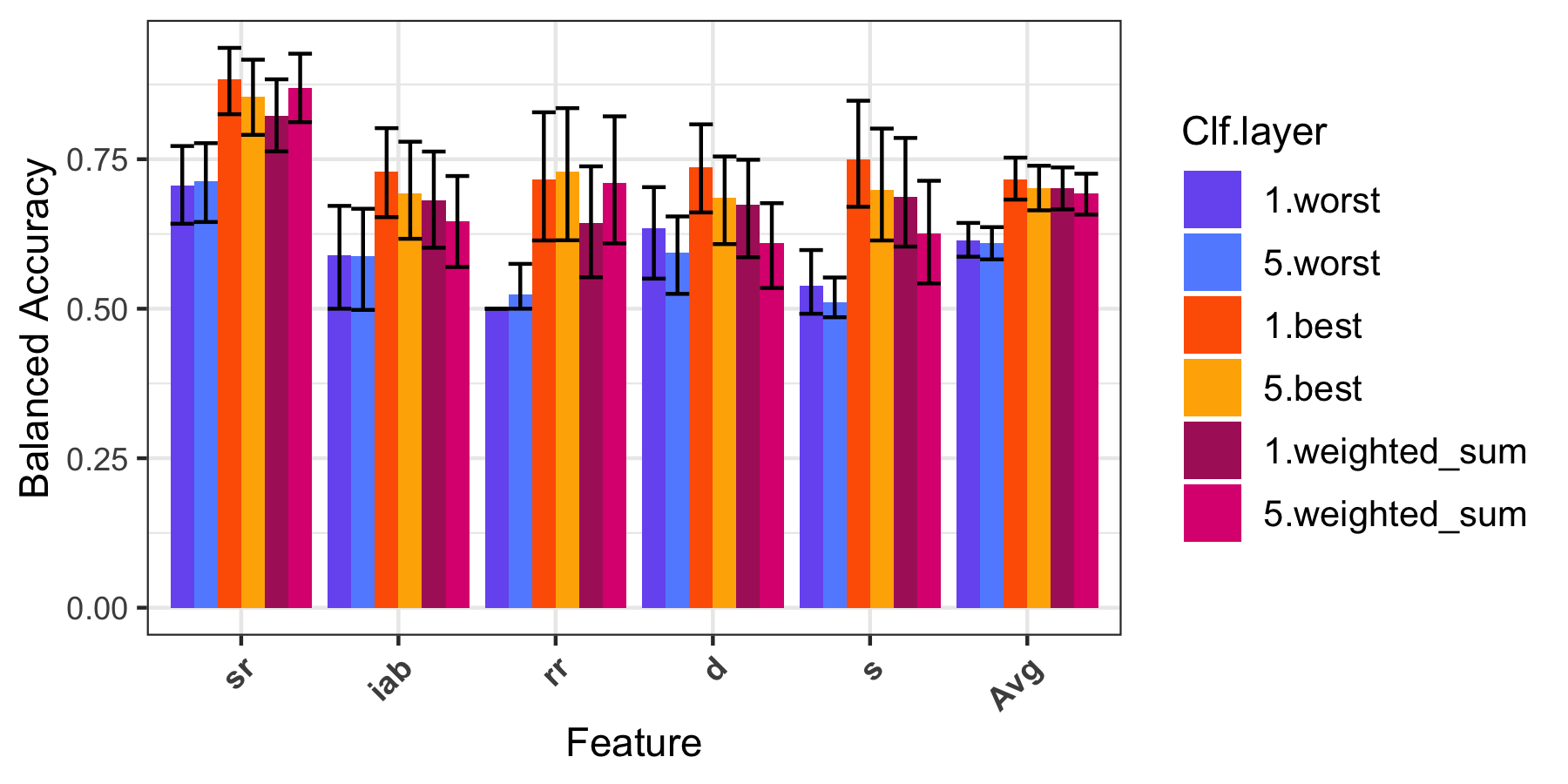}
  \caption{Comparing balanced accuracy across best and worst layers and weighted sum for each predicted pathological speech feature, with 95\% confidence intervals (AMR).}
  \label{fig:baramr}
\end{figure}

\subsection{Generalizability}
\begin{figure}[t]
  \centering
  \includegraphics[width=\linewidth]{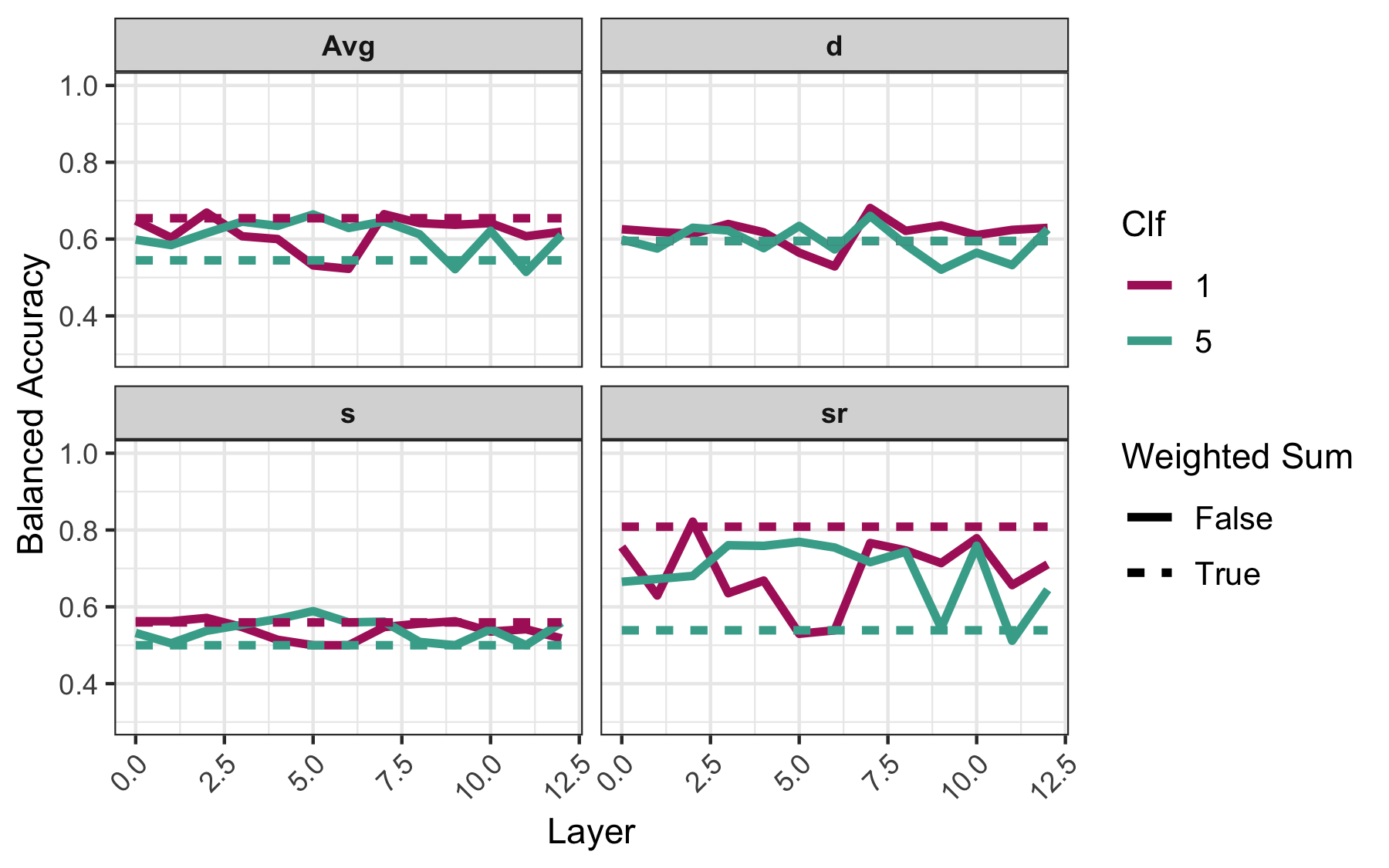}
  \caption{Comparing balanced accuracy across layers for each predicted pathological speech feature (SMR).}
  \label{fig:linesmr}
\end{figure}

Performance was worse for the out-of-distribution SMR cases and was not significantly different than the no information rate, but layer choice did have some influence on generalizability (Figure \ref{fig:linesmr}). We no longer saw a trend of performance peaking in middle layers, though earlier layers still performed better than later layers on average. This was not the case for every feature, with distortions showing later peaks. Interestingly, the best layer (5) for slow rate with AMRs (Figure \ref{fig:lineamr}) is the worst layer for slow rate with SMRs (Figure \ref{fig:linesmr}). In general, choosing the best layer resulted in an average balanced accuracy increase of $\sim$16.7\% per feature as compared to the worst layer and $\sim$6.7\% as compared to the final layer. The single classifier architecture had slightly more improvement (+1\%) based on layer choice. The weighted sum only resulted in an average increase of $\sim$8.4\% as compared to the worst layer and a decrease of $\sim$1.5\% compared to the last layer, but this was skewed by the multiple classifier weighted sum. The single classifier weighted sum notably had near best performance in the out-of-distribution tests at only $\sim$1.2\% lower than the single best layer for both slow rate and strained ($-$8.6\% for distortions) and 1.5\% lower than the average best layer (layer 2, Figure \ref{fig:linesmr}).

\section{Discussion}
We found that layer selection offers large performance improvements as compared to hyperparameter tuning or changes to classifier architecture when leveraging a speech foundation model for the downstream clinical task of predicting pathological speech features. In general, earlier layers outperformed the final layer of the model. Adding a learnable weighted sum to combine all layer representations offered comparable performance to the best average performing layer and improved upon using the final layer of the model, though it was not the most optimal. These trends did not hold for out-of-distribution tests. The best in-distribution layers were rarely the best performing layers out-of-distribution, while the learnable weighted sum almost always matched the best single layer performance. This suggests that optimizing by layer has an increased risk of overfitting. In a clinical setting, this is important to recognize as there are frequent out-of-distribution cases, both in terms of the form of the speech (a variety of elicited speech tasks) and the pathological features present in the speech. As such, while the best layer performance was consistently higher in-distribution, the generalization power of the learned weighted sum may be better suited for the demands of clinical tasks. 

Overall, our models did not perform to acceptable clinical standards, but performance was comparable to previous work. The best average in-distribution performance was 71.66\% balanced accuracy (79.3\% accuracy), while best feature performance ranged from 72.9\% (90.4\% accuracy) for rapid rate to 88.4\% (88.2\% accuracy) for slow rate in-distribution. Best out-of-distribution performance had an average balanced accuracy of 69.6\% (78\% accuracy), with best feature performance ranging from 61.7\% (79.8\% accuracy) for strained to 84\% (78.7\% accuracy) for slow rate. In comparison, larger models predicting more features with more data (multiple tasks in the input) reached 83.1\% average accuracy at the best layer \cite{Soltau23}.

Our results were also line with previous work on layer-analysis which suggests that certain representations may be better for clinical tasks \cite{Pasad23, Wagner23}. While we did not heavily explore the impact of feature, the range in performance across features may be explained by these previous studies. Given that each pathological speech feature has varied influence on a speech signal \cite{Duffy19}, a single layer may not capture all the relevant information such that the best layers vary per feature and single-feature performance drops for the average best layer. Additionally, since dataset-specific information may be encoded in representations, it follows that the best layer may be a dataset-specific parameter. As a result, the learnable weighted sum could be a valuable tool for generalization. Future work may explore the effect of features in-depth or look to improve the weighted sum, potentially by replacing it with a small network that allows the final representation to take the best parts of each layer. 

Notably, the amount of data we used as compared to the number of parameters may have led to poor learning. Our models also may not generalize to other races, ethnicities, or ages given the homogeneous nature of our dataset. Model performance and generalizability may therefore be improved by the quantity and quality of the data alone. There is also the potential that other foundation models will learn more informative representations for clinical tasks \cite{Pasad23, Wagner23, Li22}, though there is some evidence that model-internal optimization has a larger impact than model choice \cite{Soltau23}. Given the effect of layer selection shown in our results, our work suggests that continued exploration of layer selection with more data and more diverse data across a variety of foundation models will be useful for creating speech models that can be used in clinical practice. 

\section{Acknowledgements}
This work was supported by the NIH (R01 AG 83832).

\bibliographystyle{IEEEtran}
\bibliography{w2v2layers}

\end{document}